\RequirePackage{lineno} 
\documentclass[aps,prl,twocolumn,superscriptaddress,showpacs]{revtex4}
\usepackage{epsfig}
\usepackage{graphicx}
\usepackage{dcolumn}
\usepackage{amsmath}
\usepackage{xspace}
\def \Et {{\rm E}_{\rm T}}
\newcommand{\pt}{{\rm p}_{\rm T}}
\newcommand{\met}{\mbox{${\rm \not\! E}_{\rm T}$}}
\def \Ht {{\rm H}_{\rm T}}
\newcommand{\metvec}{{\not\!\! \vec{E}_T}}

\def\lgmet{\ell\gamma\met}
\def\lgmetb{\ell\gamma b \met }
\def\lgal{\ell\gamma}
\def\eemet{ee\met}

\def\Wbbbar{W\bbbar}
\def\Wccbar{W\ccbar}

\def\Wg{W\gamma}
\def\lemetb{\ell e\met b}
\def\Ztautaug{Z(\tau\tau)\gamma}
\newcommand{\pb}{{\rm pb}}
\newcommand{\invfb}{\rm fb^{-1}}
\newcommand{\ppbar}{p{\bar p}} 
\def\ccbar{c{\bar c}}
\def\bbbar{b{\bar b}}
\def\ttbar{t\bar{t}}
\def\ttg{t\bar{t}\gamma}
\newcommand{\goes}{\kern -0.08em\rightarrow\kern -0.08em}
\newcommand{\plus}{\kern -0.08em +\kern -0.08em}
\def\lplm{\ell^+\ell^-}
\def\roots{{\sqrt s}}
\newcommand{\pmasym}[2]{^{+#1}_{-#2}}

\newcommand{\Zmumu}{\mbox{Z$^0 \rightarrow \mu ^+ \mu ^-$}}
\newcommand{\Zee}{\mbox{Z$^0 \rightarrow e^+e^-$}}
\newcommand{\degs}{\mbox{$^{\circ}$}}
\newcommand{\GeV}{\ensuremath{\mathrm{Ge\kern -0.1em V}}\xspace}
\newcommand{\TeV}{\ensuremath{\mathrm{Te\kern -0.1em V}}\xspace}

\newcommand{\rrr}{\rightarrow}
\newcommand{\Etgamma}{\ensuremath{\mathrm{E_T^{\gamma}}}}
\newcommand{\Etjet}{\ensuremath{\mathrm{E_T^{jet}}}}
\newcommand{\Etelectron}{\ensuremath{\mathrm{E_T^{e}}}}
\newcommand{\Ptmuon}{\ensuremath{\mathrm{P_T^{\mu}}}}
\begin{document}
\title
{Searching the  Inclusive $\ell\gamma\met$+b-quark Signature for Radiative Top
Quark Decay and Non-Standard-Model Processes} 
\affiliation{Institute of Physics, Academia Sinica, Taipei, Taiwan 11529, Republic of China} 
\affiliation{Argonne National Laboratory, Argonne, Illinois 60439} 
\affiliation{University of Athens, 157 71 Athens, Greece} 
\affiliation{Institut de Fisica d'Altes Energies, Universitat Autonoma de Barcelona, E-08193, Bellaterra (Barcelona), Spain} 
\affiliation{Baylor University, Waco, Texas  76798} 
\affiliation{Istituto Nazionale di Fisica Nucleare Bologna, $^y$University of Bologna, I-40127 Bologna, Italy} 
\affiliation{Brandeis University, Waltham, Massachusetts 02254} 
\affiliation{University of California, Davis, Davis, California  95616} 
\affiliation{University of California, Los Angeles, Los Angeles, California  90024} 
\affiliation{University of California, San Diego, La Jolla, California  92093} 
\affiliation{University of California, Santa Barbara, Santa Barbara, California 93106} 
\affiliation{Instituto de Fisica de Cantabria, CSIC-University of Cantabria, 39005 Santander, Spain} 
\affiliation{Carnegie Mellon University, Pittsburgh, PA  15213} 
\affiliation{Enrico Fermi Institute, University of Chicago, Chicago, Illinois 60637}
\affiliation{Comenius University, 842 48 Bratislava, Slovakia; Institute of Experimental Physics, 040 01 Kosice, Slovakia} 
\affiliation{Joint Institute for Nuclear Research, RU-141980 Dubna, Russia} 
\affiliation{Duke University, Durham, North Carolina  27708} 
\affiliation{Fermi National Accelerator Laboratory, Batavia, Illinois 60510} 
\affiliation{University of Florida, Gainesville, Florida  32611} 
\affiliation{Laboratori Nazionali di Frascati, Istituto Nazionale di Fisica Nucleare, I-00044 Frascati, Italy} 
\affiliation{University of Geneva, CH-1211 Geneva 4, Switzerland} 
\affiliation{Glasgow University, Glasgow G12 8QQ, United Kingdom} 
\affiliation{Harvard University, Cambridge, Massachusetts 02138} 
\affiliation{Division of High Energy Physics, Department of Physics, University of Helsinki and Helsinki Institute of Physics, FIN-00014, Helsinki, Finland} 
\affiliation{University of Illinois, Urbana, Illinois 61801} 
\affiliation{The Johns Hopkins University, Baltimore, Maryland 21218} 
\affiliation{Institut f\"{u}r Experimentelle Kernphysik, Universit\"{a}t Karlsruhe, 76128 Karlsruhe, Germany} 
\affiliation{Center for High Energy Physics: Kyungpook National University, Daegu 702-701, Korea; Seoul National University, Seoul 151-742, Korea; Sungkyunkwan University, Suwon 440-746, Korea; Korea Institute of Science and Technology Information, Daejeon, 305-806, Korea; Chonnam National University, Gwangju, 500-757, Korea} 
\affiliation{Ernest Orlando Lawrence Berkeley National Laboratory, Berkeley, California 94720} 
\affiliation{University of Liverpool, Liverpool L69 7ZE, United Kingdom} 
\affiliation{University College London, London WC1E 6BT, United Kingdom} 
\affiliation{Centro de Investigaciones Energeticas Medioambientales y Tecnologicas, E-28040 Madrid, Spain} 
\affiliation{Massachusetts Institute of Technology, Cambridge, Massachusetts  02139} 
\affiliation{Institute of Particle Physics: McGill University, Montr\'{e}al, Qu\'{e}bec, Canada H3A~2T8; Simon Fraser University, Burnaby, British Columbia, Canada V5A~1S6; University of Toronto, Toronto, Ontario, Canada M5S~1A7; and TRIUMF, Vancouver, British Columbia, Canada V6T~2A3} 
\affiliation{University of Michigan, Ann Arbor, Michigan 48109} 
\affiliation{Michigan State University, East Lansing, Michigan  48824}
\affiliation{Institution for Theoretical and Experimental Physics, ITEP, Moscow 117259, Russia} 
\affiliation{University of New Mexico, Albuquerque, New Mexico 87131} 
\affiliation{Northwestern University, Evanston, Illinois  60208} 
\affiliation{The Ohio State University, Columbus, Ohio  43210} 
\affiliation{Okayama University, Okayama 700-8530, Japan} 
\affiliation{Osaka City University, Osaka 588, Japan} 
\affiliation{University of Oxford, Oxford OX1 3RH, United Kingdom} 
\affiliation{Istituto Nazionale di Fisica Nucleare, Sezione di Padova-Trento, $^z$University of Padova, I-35131 Padova, Italy} 
\affiliation{LPNHE, Universite Pierre et Marie Curie/IN2P3-CNRS, UMR7585, Paris, F-75252 France} 
\affiliation{University of Pennsylvania, Philadelphia, Pennsylvania 19104}
\affiliation{Istituto Nazionale di Fisica Nucleare Pisa, $^{aa}$University of Pisa, $^{bb}$University of Siena and $^{cc}$Scuola Normale Superiore, I-56127 Pisa, Italy} 
\affiliation{University of Pittsburgh, Pittsburgh, Pennsylvania 15260} 
\affiliation{Purdue University, West Lafayette, Indiana 47907} 
\affiliation{University of Rochester, Rochester, New York 14627} 
\affiliation{The Rockefeller University, New York, New York 10021} 
\affiliation{Istituto Nazionale di Fisica Nucleare, Sezione di Roma 1, $^{dd}$Sapienza Universit\`{a} di Roma, I-00185 Roma, Italy} 

\affiliation{Rutgers University, Piscataway, New Jersey 08855} 
\affiliation{Texas A\&M University, College Station, Texas 77843} 
\affiliation{Istituto Nazionale di Fisica Nucleare Trieste/Udine, I-34100 Trieste, $^{ee}$University of Trieste/Udine, I-33100 Udine, Italy} 
\affiliation{University of Tsukuba, Tsukuba, Ibaraki 305, Japan} 
\affiliation{Tufts University, Medford, Massachusetts 02155} 
\affiliation{Waseda University, Tokyo 169, Japan} 
\affiliation{Wayne State University, Detroit, Michigan  48201} 
\affiliation{University of Wisconsin, Madison, Wisconsin 53706} 
\affiliation{Yale University, New Haven, Connecticut 06520} 
\author{T.~Aaltonen}
\affiliation{Division of High Energy Physics, Department of Physics, University of Helsinki and Helsinki Institute of Physics, FIN-00014, Helsinki, Finland}
\author{J.~Adelman}
\affiliation{Enrico Fermi Institute, University of Chicago, Chicago, Illinois 60637}
\author{T.~Akimoto}
\affiliation{University of Tsukuba, Tsukuba, Ibaraki 305, Japan}
\author{B.~\'{A}lvarez~Gonz\'{a}lez$^t$}
\affiliation{Instituto de Fisica de Cantabria, CSIC-University of Cantabria, 39005 Santander, Spain}
\author{S.~Amerio$^z$}
\affiliation{Istituto Nazionale di Fisica Nucleare, Sezione di Padova-Trento, $^z$University of Padova, I-35131 Padova, Italy} 

\author{D.~Amidei}
\affiliation{University of Michigan, Ann Arbor, Michigan 48109}
\author{A.~Anastassov}
\affiliation{Northwestern University, Evanston, Illinois  60208}
\author{A.~Annovi}
\affiliation{Laboratori Nazionali di Frascati, Istituto Nazionale di Fisica Nucleare, I-00044 Frascati, Italy}
\author{J.~Antos}
\affiliation{Comenius University, 842 48 Bratislava, Slovakia; Institute of Experimental Physics, 040 01 Kosice, Slovakia}
\author{G.~Apollinari}
\affiliation{Fermi National Accelerator Laboratory, Batavia, Illinois 60510}
% ...

\collaboration{CDF Collaboration\footnote{With visitors from $^a$University of Massachusetts Amherst, Amherst, Massachusetts 01003,
$^b$Universiteit Antwerpen, B-2610 Antwerp, Belgium, 
$^c$University of Bristol, Bristol BS8 1TL, United Kingdom,
$^d$Chinese Academy of Sciences, Beijing 100864, China, 
$^e$Istituto Nazionale di Fisica Nucleare, Sezione di Cagliari, 09042 Monserrato (Cagliari), Italy,
$^f$University of California Irvine, Irvine, CA  92697, 
$^g$University of California Santa Cruz, Santa Cruz, CA  95064, 
$^h$Cornell University, Ithaca, NY  14853, 
$^i$University of Cyprus, Nicosia CY-1678, Cyprus, 
$^j$University College Dublin, Dublin 4, Ireland,
$^k$University of Edinburgh, Edinburgh EH9 3JZ, United Kingdom, 
$^l$University of Fukui, Fukui City, Fukui Prefecture, Japan 910-0017
$^m$Kinki University, Higashi-Osaka City, Japan 577-8502
$^n$Universidad Iberoamericana, Mexico D.F., Mexico,
$^o$University of Iowa, Iowa City, IA  52242,
$^p$Queen Mary, University of London, London, E1 4NS, England,
$^q$University of Manchester, Manchester M13 9PL, England, 
$^r$Nagasaki Institute of Applied Science, Nagasaki, Japan, 
$^s$University of Notre Dame, Notre Dame, IN 46556,
$^t$University de Oviedo, E-33007 Oviedo, Spain, 
$^u$Texas Tech University, Lubbock, TX  79609, 
$^v$IFIC(CSIC-Universitat de Valencia), 46071 Valencia, Spain,
$^w$University of Virginia, Charlottesville, VA  22904,
$^x$Bergische Universit\"at Wuppertal, 42097 Wuppertal, Germany,
$^{ff}$On leave from J.~Stefan Institute, Ljubljana, Slovenia, 
}}
\noaffiliation

\date{\today}

\begin{abstract}

In a search for new phenomena in a signature suppressed in the
standard model of elementary particles (SM), we compare the inclusive
production of events containing a lepton ($\ell$), a photon
($\gamma$), significant transverse momentum imbalance ($\met$), and a
jet identified as containing a b-quark, to SM predictions.  The search
uses data produced in proton-antiproton collisions at $\sqrt{s} =
1.96~\TeV$ corresponding to 1.9 $\invfb$ of integrated luminosity
taken with the CDF detector at the Fermilab Tevatron. We find $28$
$\lgmetb$ events versus an expectation of $31.0\pmasym{4.1}{3.5}$
events.  If we further require events to contain at least three jets
and large total transverse energy, simulations predict that the
largest SM source is top-quark pair production with an additional
radiated photon, $t\bar{t}+\gamma$.  In the data we observe $16$
$\ttg$ candidate events versus an expectation from SM sources of $11.2
\pmasym{2.3}{2.1}$.  Assuming the difference between the observed
number and the predicted non-top-quark total is due to SM top quark
production, we estimate the $\ttg$ cross section to be $0.15
\pm 0.08~\pb$.

\end{abstract}


\maketitle

The unknown nature of possible new phenomena in the energy range
accessible at the Tevatron collider is the motivation for a search
strategy~\cite{Toback_all, Berryhill_all, Knuteson} that does not
focus on current hypothetical models of new physics, but instead tests
the standard model (SM)~\cite{SM}. The emphasis on presenting
measurements and SM predictions, rather than comparisons with
arbitrarily chosen other models, allows a wide net for physics beyond
the SM that can be used now by proponents of current models of `new
physics' as well as in the future by theorists with new ideas and
facts.  Here we report the results of a search for events containing a
lepton ($\ell$), a photon ($\gamma$), significant transverse momentum
imbalance ($\met$)~\cite{EtPt}, and a jet identified as containing a
b-quark; i.e. the final state $\lgmetb+X$. This channel, which
contains a vector boson and a third-generation quark, is suppressed in
the SM, and is consequently sensitive to rare new phenomena.  The data
correspond to an integrated luminosity of $1.9$ $\invfb$ of $\ppbar$
collisions at $\roots = 1.96~\TeV$, collected using the CDF II
detector~\cite{CDFII}.  This search is an extension of a previous
search in the lepton+photon+X signature, described in detail in
Ref.~\cite{Loginov_all}.

A search for SM production of top-quark pairs with an additional
photon, $\ttg$, is a natural extension of the $\lgmetb+X$ analysis.
By further requiring events to contain at least three jets and large
total transverse energy ($\Ht$)~\cite{EtPt}, we find that the SM
predicts the largest source of events will be top-quark pair
production with an additional radiated photon, $t\bar{t}+\gamma$. The
process is of interest for the direct measurement of the electric
charge of the top quark~\cite{Baur_all}, as well as being another
low-cross-section search signature in which rare non-SM processes
could appear.


The CDF II detector~\cite{CDFII} is a cylindrically-symmetric magnetic
spectrometer designed to study $\ppbar$ collisions at the Fermilab
Tevatron. Here we briefly describe the detector
subsystems relevant for the present analysis.

Tracking systems are used to measure the momenta of charged particles
and to identify leptons with large transverse
momenta~\cite{EtPt}. A multi-layer system of silicon strip
detectors~\cite{SVX}, which identifies tracks in both the $r$-$\phi$
and $r$-$z$ views~\cite{CDF_coo}, and the central outer tracker
(COT)~\cite{COT}, are contained in a
superconducting solenoid that generates a magnetic field of 1.4 T. The
COT is a 3.1 m long open-cell drift chamber that makes up to 96
measurements along the track of each charged particle in the region
$|\eta|<1$. Sense wires are arranged in 8 alternating axial and $\pm
2\degs$ stereo superlayers with 12 wire layers each. For high-momentum
tracks, the COT transverse momentum~($\pt$) resolution is
$\sigma_{\pt}/\pt^2 \simeq 0.0017~\GeV^{-1}$~\cite{COT}.

Segmented calorimeters with towers arranged in a projective geometry,
each tower consisting of an electromagnetic and an hadronic
compartment~\cite{cem_resolution, cal_upgrade}, cover the region
$|\eta|<3.6$.  In this analysis we select photons and electrons in the
central region, $|\eta|<1$, where a system (CES) with finer spatial
resolution is used to make profile measurements of electromagnetic
showers at shower maximum~\cite{CDFII}. Electrons are reconstructed in
the central electromagnetic calorimeter (CEM) with an $\Et$ resolution
of $\sigma(\Et)/\Et \simeq 13.5\%/\sqrt{\Et/\GeV}\oplus
2\%$~\cite{cem_resolution}.
Jets are identified in the electromagnetic and hadronic calorimeters
using a cone in $\eta-\phi$ space of radius
0.4~\cite{Blazey:2000qt,jet_counting}.  The jet energy resolution is
approximately $\sigma \simeq 0.1\times\Et~(\GeV)~
+1.0~\GeV$~\cite{jet_resolution}.

Muons are identified using the central muon (CMU), the central muon
upgrade (CMP), and the central muon extension (CMX)
systems~\cite{cmu_ref, cmp_ref}, which cover the kinematic region
$|\eta|<1$. The CMU uses four layers of planar drift chambers to
detect muons with $\pt>1.4~\GeV$ in the region of $|\eta|<0.6$. The
CMP consists of four additional layers of planar drift chambers
located behind 0.6~m of steel outside the magnetic return yoke, and
detects muons with $\pt>2.0~\GeV$. The CMX detects muons in the region
$0.6<|\eta|<1.0$ with four to eight layers of drift chambers,
depending on the polar angle.

We use identification algorithms that exploit the long lifetime
($c\tau_0 \sim 450~\mu m$) of $b$ hadrons to identify jets containing
$b$ hadrons. Candidate $b$-jets are identified through the presence of
a secondary decay vertex displaced from the beam line in the region
$|\eta| < 2$~\cite{secvtx}.

The beam luminosity is measured using two arrays of gas Cherenkov
counters, located in the region $3.7<|\eta|<4.7$. The total
uncertainty on the luminosity has been estimated to be 6\%, where 4.4\%
comes from the acceptance and operation of the luminosity monitor and
4.0\% from the calculation of the accepted inelastic $\ppbar$
cross section~\cite{luminosity}.

We use events selected by the online event selection (trigger)
system~\cite{CDFII} to have a high $\pt$ electron or muon in the
central region, $|\eta| \lesssim 1.0$. The electron trigger requires a
cluster of energy in the central electromagnetic calorimeter with a
COT track pointing at the cluster. The muon trigger requires a COT
track that extrapolates to a track segment in the muon chambers.

Inclusive $\lgal$ events are selected by requiring a central
high-energy $\gamma$ candidate and a central high-energy $e$ or $\mu$
candidate originating less than 60 cm along the beam-line from the
detector center and passing the selection criteria listed below.  To
reduce background from the decays of hadrons produced in jets, both
the photon and the lepton in each event are required to be
isolated~\cite{PhoIso}.

An electron candidate must meet the following selection criteria: a) a
high-quality track \cite{ElePt} with $\pt>0.5~\Et$, unless $\Et >
100~\GeV$, in which case the $\pt$ threshold is set to $20~\GeV$; b) a
good transverse shower profile that matches the extrapolated track
position; c) a lateral sharing of energy in the calorimeter towers
containing the electron shower consistent with that expected; and d)
minimal leakage into the hadron calorimeter~\cite{hadoem}.

A muon candidate must have: a) a well-measured track in the COT; b)
energy deposited in the calorimeter consistent with expectations; c) a
muon track segment in both the CMU and CMP, or in the CMX, consistent
with the extrapolated COT track; and d) COT timing consistent with a
track from a $\ppbar$ collision.

Photon candidates are required to have no track with $\pt>1~\GeV$, and
at most one track with $\pt<1~\GeV$, pointing at the calorimeter
cluster, good profiles in both transverse dimensions at shower
maximum, and minimal leakage into the hadron
calorimeter~\cite{hadoem}.

\begin{figure}[!b]
\vspace*{-0.2in}
\begin{center}
\hspace*{-0.1in}
\includegraphics*[angle =90,width=0.50\textwidth]{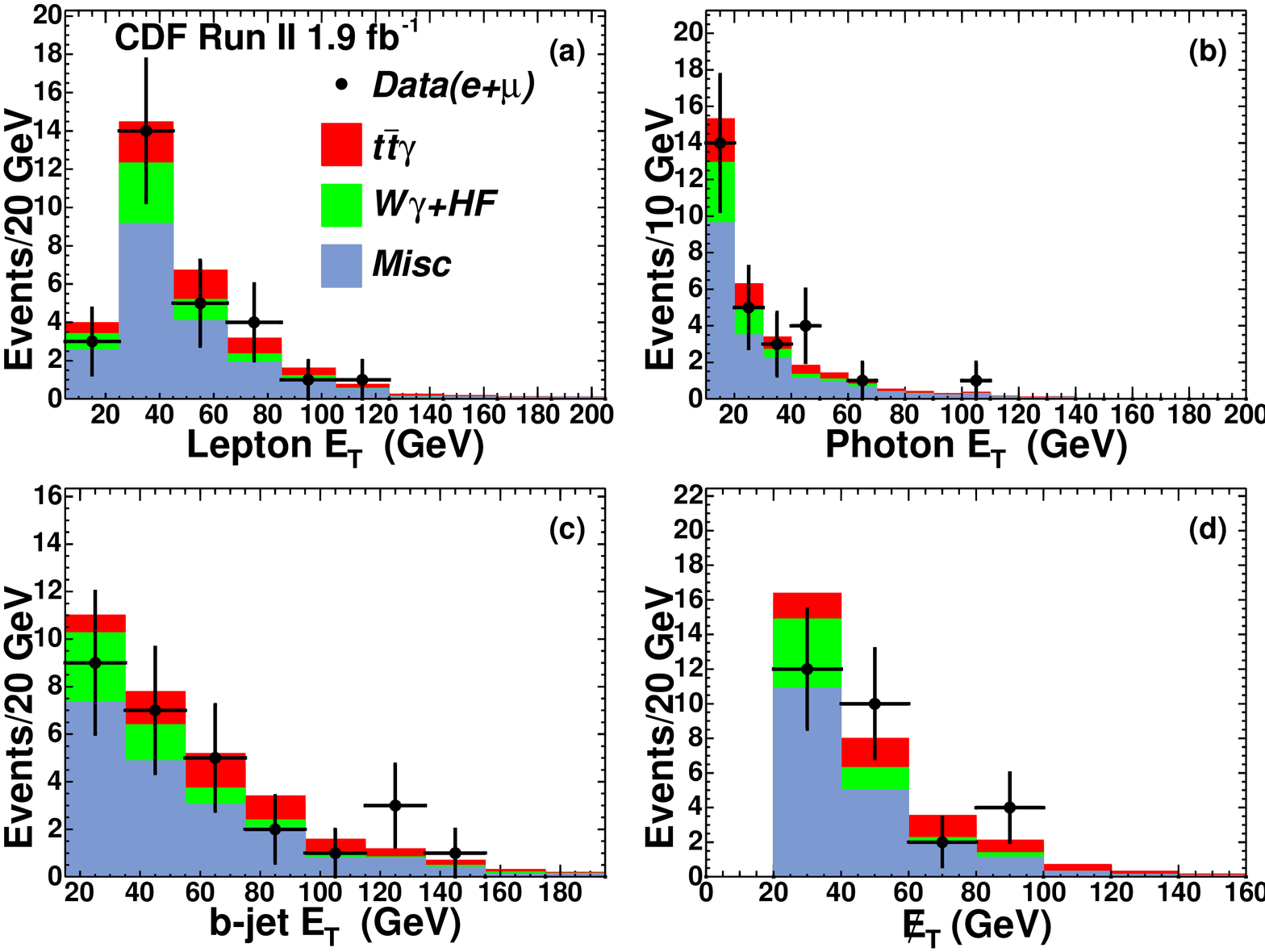}
\end{center}
\vspace*{-0.2in}
\caption{ The distributions for events in the $\lgmetb$ sample (points)
  in a) the $\Et$ of the lepton; b) the $\Et$ of the photon; c) the
  $\Et$ of the most energetic $b$-jet in an event; and d) the missing
  transverse energy. The histograms show the estimated SM
  contributions from radiative top quark decay($\ttg$), 
  WZ production, W$\gamma$ production with heavy flavor (HF),
$\tau$ leptons, electrons, and jets misidentified as
  photons, mistagged light-quark and gluon jets, and jets
  misidentified as leptons (QCD). }
\label{figure1}
\end{figure}

Missing transverse energy ($\met$) is calculated from the calorimeter
tower energies in the region $|\eta| < 3.6$. Corrections are then made
to the $\met$ for the position of the reconstructed primary vertex,
and for non-uniform calorimeter response~\cite{jet_corr} for jets with
uncorrected $\Et > 15~\GeV$ and $|\eta| < 2.0$, and for muons with
$\Ptmuon > 20~\GeV$.

\begin{figure}[!t]
\vspace*{-0.2in}
\begin{center}
\hspace*{-0.1in}
\includegraphics*[angle=90,width=0.50\textwidth]{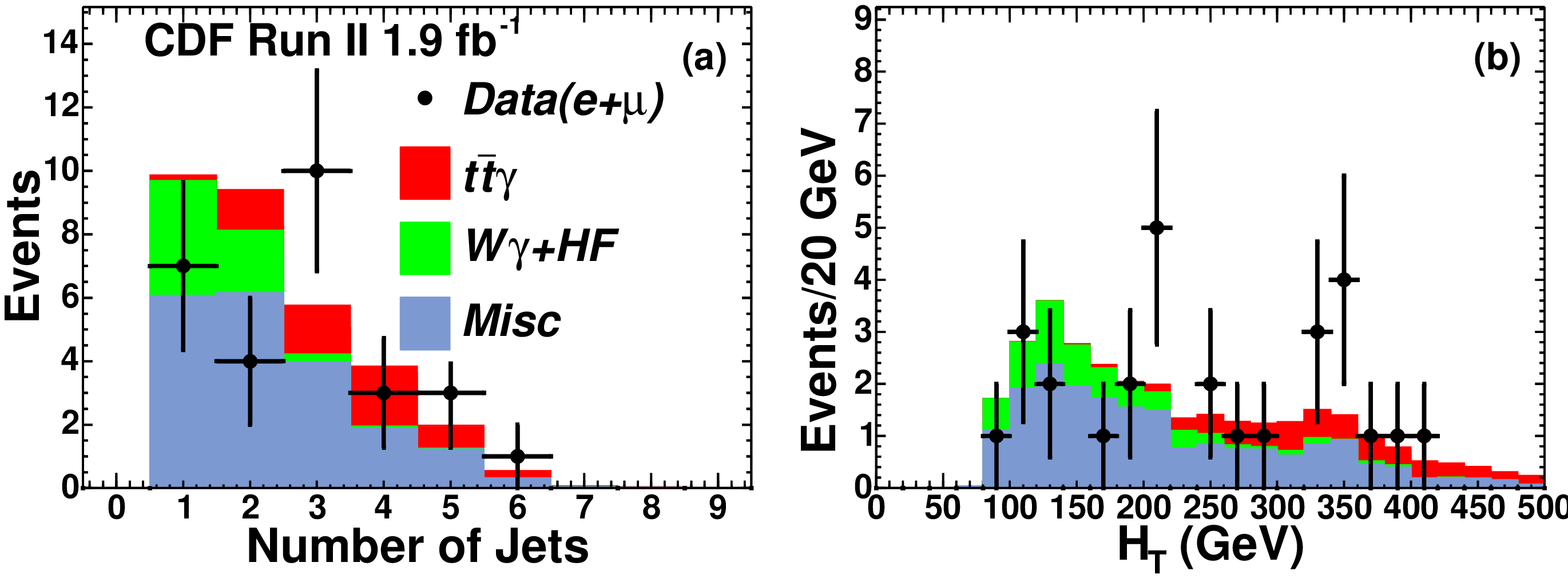}
\end{center}
\vspace*{-0.2in}
\caption
{The distributions for events in the $\lgmetb$ sample (points) in a)
the total number of jets; b) the total transverse energy $\Ht$ for the
$\lgmetb$ events. The histograms show the estimated SM contributions
from radiative top quark decay($\ttg$), WZ production, W$\gamma$
production with heavy flavor (HF), $\tau$ leptons, electrons, and jets
misidentified as photons, mistagged light-quark and gluon jets, and
jets misidentified as leptons (QCD). }
\label{figure2}
\end{figure}

The inclusive $\lgmetb$ search is defined by requiring that an event
contain a central electron (or muon) with $\Etelectron
(\Ptmuon)>20~\GeV$ ~\cite{EtPt}, a central photon with
$\Etgamma>10~\GeV$, a $b$-tagged jet with $\Etjet>15~\GeV$, and $\met>
20~\GeV$~\cite{baur}. Figures~\ref{figure1} and
\ref{figure2} show kinematic distributions for
events in the $\lgmetb$ sample.  

\begin{table}[!t]
\begin{center}
\vspace*{-0.2in}
\caption
{Summary for the $\lgmetb$ search. Backgrounds from
$WW$, $ZZ$, and single top quark with an additional radiated photon
are found to be negligible.}
\label{table1}
\begin{tabular}{l c c c}
\hline\hline
\multicolumn{4}{c} {
Lepton + Photon + $\met$ + $b$ Events}\\
\hline
SM Source & $e\gamma b\met$ & $\mu\gamma b\met$ & $(e+\mu)\gamma b\met$ \\
\hline
$\ttg$ semileptonic                                        & 
$2.06  \pm 0.38$   &
$1.52 \pm 0.28$  &
$3.58  \pm 0.65$   \\
$\ttg$ dileptonic                                        & 
$1.30  \pm 0.23$   &
$1.02 \pm 0.18$  &
$2.32  \pm 0.41$   \\
$W^{\pm}c\gamma$                         & 
$1.58  \pm 0.83$     &
$1.51 \pm 0.80$    &
$3.09  \pm 1.59$     \\
$W^{\pm}cc\gamma$                         & 
$0.17  \pm 0.12  $    &
$0.46 \pm 0.26 $    &
$0.63  \pm 0.35  $     \\
$W^{\pm}bb\gamma$                         & 
$1.30  \pm 0.67  $     &
$0.88 \pm 0.46 $    &
$2.18  \pm 1.11  $     \\
$\Ztautaug$                                          & 
$0.13  \pm 0.09  $     &
$0.11 \pm 0.08 $    &
$0.24  \pm 0.12  $     \\
$WZ$                                          & 
$0.08  \pm 0.04  $     &
$0.01 \pm 0.01 $    &
$0.09  \pm 0.04  $     \\
$\tau \rrr \gamma$ fake                                        & 
$0.12  \pm 0.04  $   &
$0.10 \pm 0.03 $  &
$0.22  \pm 0.05  $   \\
\hline
Jet faking $\gamma$& 
$4.56  \pm 1.92$       &
$3.02 \pm 1.19$      &
$7.58  \pm 3.11$       \\
Mistagged $b$-jets                                       & 
$4.11  \pm 0.41$       &
$3.54 \pm 0.37$      &
$7.65  \pm 0.70$       \\
QCD     & 
$1.5  \pm 0.8 $   &
$0.0 \pmasym{1.0}{0.0}$   &
$1.5  \pmasym{1.3}{0.8} $   \\
$\eemet b$, $e\goes\gamma$                      & 
$1.50 \pm 0.28$    &
--                                       &
$1.50 \pm 0.28$    \\
$\mu e\met b$, $e\goes\gamma$                      & 
--                                       &
$0.45 \pm 0.10$    &
$0.45 \pm 0.10$    \\
\hline
Predicted                    & 
$18.4  \pm 2.4 (tot)$    &
$12.6 \pmasym{1.9}{1.6}(tot)$    &
$31.0  \pmasym{4.1}{3.9} (tot)$    \\
\hline
Observed                       & 
$16$                          & 
$12$                         & 
$28$                          \\
\hline
\hline
\end{tabular}

\vspace*{-0.2in}
\end{center}

\end{table}

\begin{figure}[!b]
\vspace*{-0.2in}
\begin{center}
\hspace*{-0.1in}
\includegraphics*[angle =90,width=0.50\textwidth]{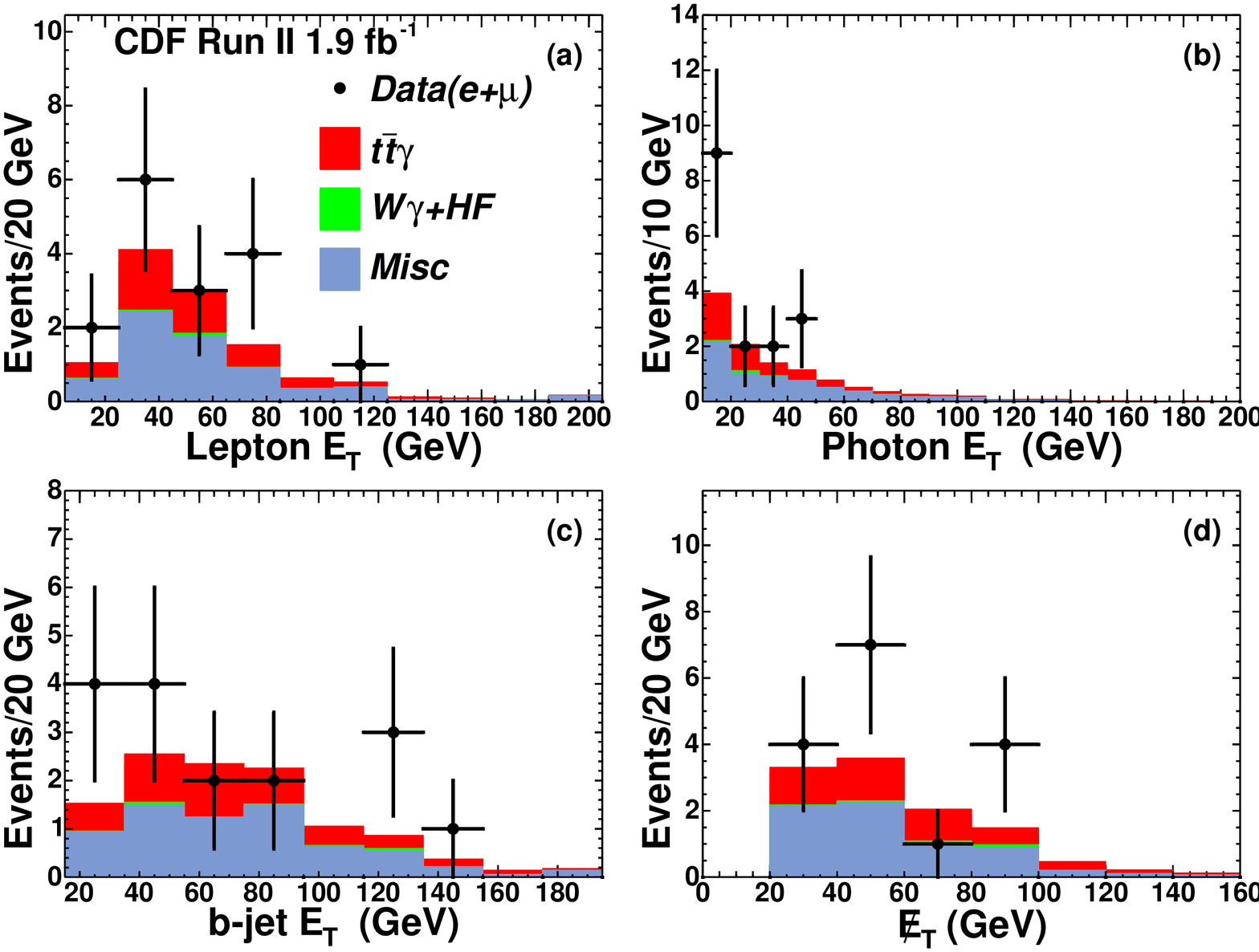}
\end{center}
\vspace*{-0.2in}
\caption{ The distributions for events in the $\ttg$ sample (points)
  in a) the $\Et$ of the lepton; b) the $\Et$ of the photon; c) the
  $\Et$ of the most energetic $b$-jet in an event; and d) the missing
  transverse energy, $\met$.  The histograms show the estimated SM
  contributions from radiative top quark decay($\ttg$), WZ production,
  W$\gamma$ production with heavy flavor (HF), $\tau$ leptons,
  electrons, and jets misidentified as photons, mistagged
  light-quark and gluon jets, and jets misidentified as leptons (QCD). }
\label{figure3}
\end{figure}

\begin{table}[!b]
\begin{center}
\caption
{Summary of the expected SM contributions to the $\ttg$
search. Backgrounds from $WW$, $ZZ$, single top quark with an
additional radiated photon are found to be negligible.}
\label{table2}
\begin{tabular}{l c c  c }
\hline\hline
\multicolumn{4}{c} {
$\ttg$}\\
\hline
SM Source & $e\gamma b\met$ & $\mu\gamma b\met$ & $(e+\mu)\gamma b\met$ \\
\hline
$\ttg$ semileptonic                                           & 
$1.97  \pm 0.36$   &
$1.47 \pm 0.27$  &
$3.44  \pm 0.62$   \\
$\ttg$ dileptonic                                    & 
$0.52  \pm 0.10  $   &
$0.43 \pm 0.08 $  &
$0.95  \pm 0.17  $   \\
$W^{\pm}c\gamma$                         & 
$0.0  \pmasym{0.05 }{0} $     &
$0.0 \pmasym{0.05}{0} $    &
$0  \pmasym{0.07 }{0} $     \\
$W^{\pm}cc\gamma$                         & 
$0.0  \pmasym{0.04}{0}  $     &
$0.03 \pm 0.03 $    &
$0.03  \pmasym{0.05}{0.03}  $     \\
$W^{\pm}bb\gamma$                         & 
$0.13  \pm 0.08  $     &
$0.02 \pm 0.02 $    &
$0.15  \pm 0.09  $     \\
$WZ$                                          & 
$0.02  \pm 0.02  $     &
$0.0 \pmasym{0.02}{0} $    &
$0.02  \pm 0.02  $     \\
$\tau \rrr \gamma$ fake                                        & 
$0.08  \pm 0.01  $   &
$0.02 \pm 0.01 $  &
$0.10  \pm 0.01  $   \\
\hline
Jet faking $\gamma$& 
$2.37  \pm 1.22$       &
$1.42 \pm 0.70$      &
$3.79  \pm 1.92$       \\
Mistagged $b$-jets                                       & 
$0.78  \pm 0.20$       &
$0.83 \pm 0.22$      &
$1.61  \pm 0.31$       \\
QCD     & 
$0.5  \pm 0.5 $   &
$0.0 \pmasym{1.0}{0.0}$   &
$0.5  \pmasym{1.1}{0.5} $   \\
$\eemet b$, $e\goes\gamma$                      & 
$0.34 \pm 0.11$    &
--                                       &
$0.34 \pm 0.11$    \\
$\mu e\met b$, $e\goes\gamma$                      & 
--                                       &
$0.20 \pm 0.06$    &
$0.20 \pm 0.06$    \\
\hline
Predicted                    & 
$ 6.7  \pm 1.4 (tot)$    &
$ 4.4 \pmasym{1.3}{0.8}(tot)$    &
$ 11.2 \pmasym{2.3}{2.1} (tot)$    \\
\hline
Observed                       & 
$8$                          & 
$8$                         & 
$16$                          \\
\hline
\hline
\end{tabular}

\vspace*{-0.2in}
\end{center}

\end{table}

The dominant SM sources of $\lgmetb$ events at the Tevatron are $\ttg$
production and $W\gamma$+heavy flavour (HF) ($Wc\gamma$,
$\Wccbar\gamma$, $\Wbbbar\gamma$) in which a $W$ boson decays
leptonically ($\ell \nu$) and a photon is radiated from an
initial-state quark, the $W$, or a charged final-state
lepton~\cite{CDF_WZgamma}.  We use the { \sc madgraph} matrix-element
event generator~\cite{MadGraph} to estimate these contributions.
Initial state radiation and parton showering are simulated by the {\sc
pythia} shower code~\cite{Pythia} tuned to reproduce the underlying
event~\cite{Tune_A}. The generated particles are then passed through a
full simulation of the detector, and these events are then
reconstructed with the same reconstruction code used for the data.
The expected contributions from $\ttg$ and $\Wg+HF$ production to the
$\lgmetb$ and $\ttg$ searches are given in Tables~\ref{table1}
and~\ref{table2}. A correction for higher-order processes (K-factor)
of 1.10 $\pm$ 0.15 for the $\ttg$~\cite{baur} has been applied to the
LO MC estimates. We have also applied a K-factor of 2.10 $\pm$ 1.05
for the $\Wg+HF$~\cite{K-HF}. Backgrounds from $WW$, $ZZ$, and the
production of a single top quark plus a photon are estimated to be
negligible.


The background from top decays in which tau leptons are misidentified
as photons is estimated from a
$\ttbar$ {\sc pythia}~\cite{Pythia} sample using simulation
information to identify tau leptons and then applying the same
analysis selection criteria as for data.

\begin{figure}[!t]
\vspace*{-0.2in}
\begin{center}
\hspace*{-0.1in}
\includegraphics*[angle=90,width=0.50\textwidth]{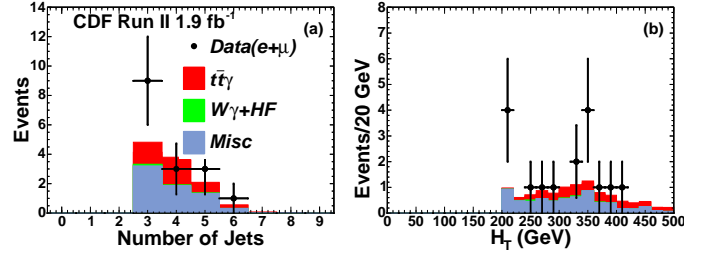}
\end{center}
\vspace*{-0.2in}
\caption
{The distributions for events in the $\ttg$ sample (points) in a) the
total number of jets; b) the total transverse energy $\Ht$ for the
$\lgmetb$ events. The histograms show the estimated SM
  contributions from radiative top quark decay($\ttg$), 
  WZ production, W$\gamma$ production with heavy flavor (HF),
$\tau$ leptons, electrons, and jets misidentified as
  photons, mistagged light-quark and gluon jets, and jets
  misidentified as leptons (QCD). }
\label{figure4}
\end{figure}

High-$\pt$ photons are copiously created from hadron decays in jets
initiated by a scattered quark or gluon. In particular, mesons such as
the $\pi^0$ or $\eta$ decay to multiple photons which may pass the
photon selection criteria. To estimate the number of events with a jet
misidentified as a photon, we first measure the probability for a jet
to be misidentified as a photon, $\mathrm{P^{jet}_{\gamma}(\Et)}$, as a
function of the measured $\Etjet$, in data samples triggered on
jets. We then measure the jet $\Et$ in $\ell\met b+$jet and $\ell\met
+ >3$ jets $(\Ht>200~\GeV)$ samples, respectively, and multiply by
$P^{jet}_{\gamma}(\Et)$. An uncertainty of 50\% on the number of such
events is calculated by using the measured jet spectrum and the upper
and lower bounds on the $\Et$-dependent misidentification
rate~\cite{CDF_WZgamma}.

To estimate the probability to mistakenly b-tag a light jet (a
mistag), each jet in the $\lgmet$+pretagged jet sample is weighted by
its mistag rate that is obtained from tagged events in which the
b-hadron decay vertex is measured to be on the opposite side of the
primary vertex from the direction of the jet, an unphysical
geometry. The mistag rate derived from these `negative' tags provides
an estimate of the number of false positive tags after a correction
for interactions in material in the inner tracking volume and
long-lived light-flavor particles. The mistag rate per jet is measured
using a large inclusive-jet data sample.

We have estimated the background due to events with jets misidentified
as high-$\pt$ leptons by studying the total $\pt$ of tracks in a cone
in $\eta-\varphi$ space of radius R=0.4 around the lepton track (track
isolation)~\cite{Loginov_all}. We compared the distribution of track
isolation in the signal sample to that of the $\Zee$ and $\Zmumu$ data
samples, and to that of the QCD background data sample, which is 
dominated by light-flavor and gluon jets.

The number of events with an electron misidentified as a photon
expected in the $\lgmetb$ sample is determined by measuring the
electron $\Et$ spectrum in $\lemetb$ samples, and then multiplying by
$P_{e\rrr\gamma}$, the probability of an electron being misidentified
as a photon. We determine $P_{e\rrr\gamma}$ from $\Zee$ events in
which one of the electrons radiates a high-$\Et$ photon, resulting in
an electron-photon system with an invariant mass consistent with that
of the Z-boson.

The uncertainties on the numbers of expected events for the $\lgmetb$
search listed in Tables~\ref{table1} and \ref{table2} include
systematic and statistical uncertainties. A total uncertainty of 6\%
is quoted for the luminosity measurements~\cite{luminosity}.
The systematics relevant to the $\lgmetb$ and $\ttg$ analyses also
include a 5\% uncertainty on the $b$-tagging efficiency, and
uncertainties on the K-factors of 15\% for the $\ttbar\gamma$ MC
samples and 50\% for the $\Wg+HF$ samples. The largest experimental
systematic uncertainty comes from the rate of misidentifying jets as
photons, which we estimate to be uncertain to approximately 50\%.

We find $28$ $\lgmetb$ events versus an expectation of $31.0
\pmasym{4.1}{3.5}$ events. The data agree well with
the SM predictions, with the precision of the comparison being limited
by statistics for the present integrated luminosity.


\begin{figure}[!b]
\vspace*{-0.2in}
\begin{center}
\hspace*{-0.1in}
\includegraphics*[angle =90,width=0.40\textwidth]{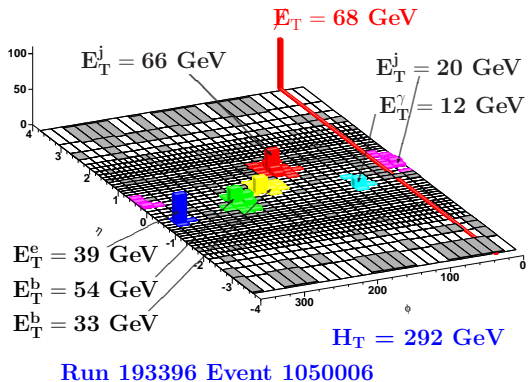}
\end{center}
\vspace*{-0.2in}
\caption{The $\eta-\phi$ plot of a $\ttg$ candidate event, in which the
  energies deposited in the calorimeter towers are displayed in the
  $\eta$-$\phi$ plane. The reconstructed top quark mass is $167~\GeV$;
  the photon $\Et$ is $12~\GeV$.}
\label{figure6}
\end{figure}

A second search, for $\ttg$ events, is constructed by further
requiring $\Ht>200~\GeV$~\cite{top_dileptonic} and $N_{jets}>2$, where
$N_{jets}$ is the number of jets in the event~\cite{jet_counting}.  We
observe $16$ $\ttg$ candidate events.
Figures~\ref{figure3} and \ref{figure4} show the corresponding
kinematic distributions for events in the $\ttg$ subsample.  An event
display of a $\ttg$ candidate event is shown in Fig.~\ref{figure6}.

For the $\ttg$ search, the detection efficiency and acceptance are
calculated using {\sc madgraph} to generate $\ttg$ events with one
leptonic $W$ decay. As in the $\lgmetb$ search, the generated
particles are then passed through a full detector simulation of the
detector and are then reconstructed with the same reconstruction code
used for the data.  We find a SM expectation of
$11.2 \pmasym{2.3}{2.1}$ events. 

\begin{figure}[!t]
\vspace*{-0.2in}
\begin{center}
\hspace*{-0.1in}
\includegraphics*[angle =0,width=0.40\textwidth]{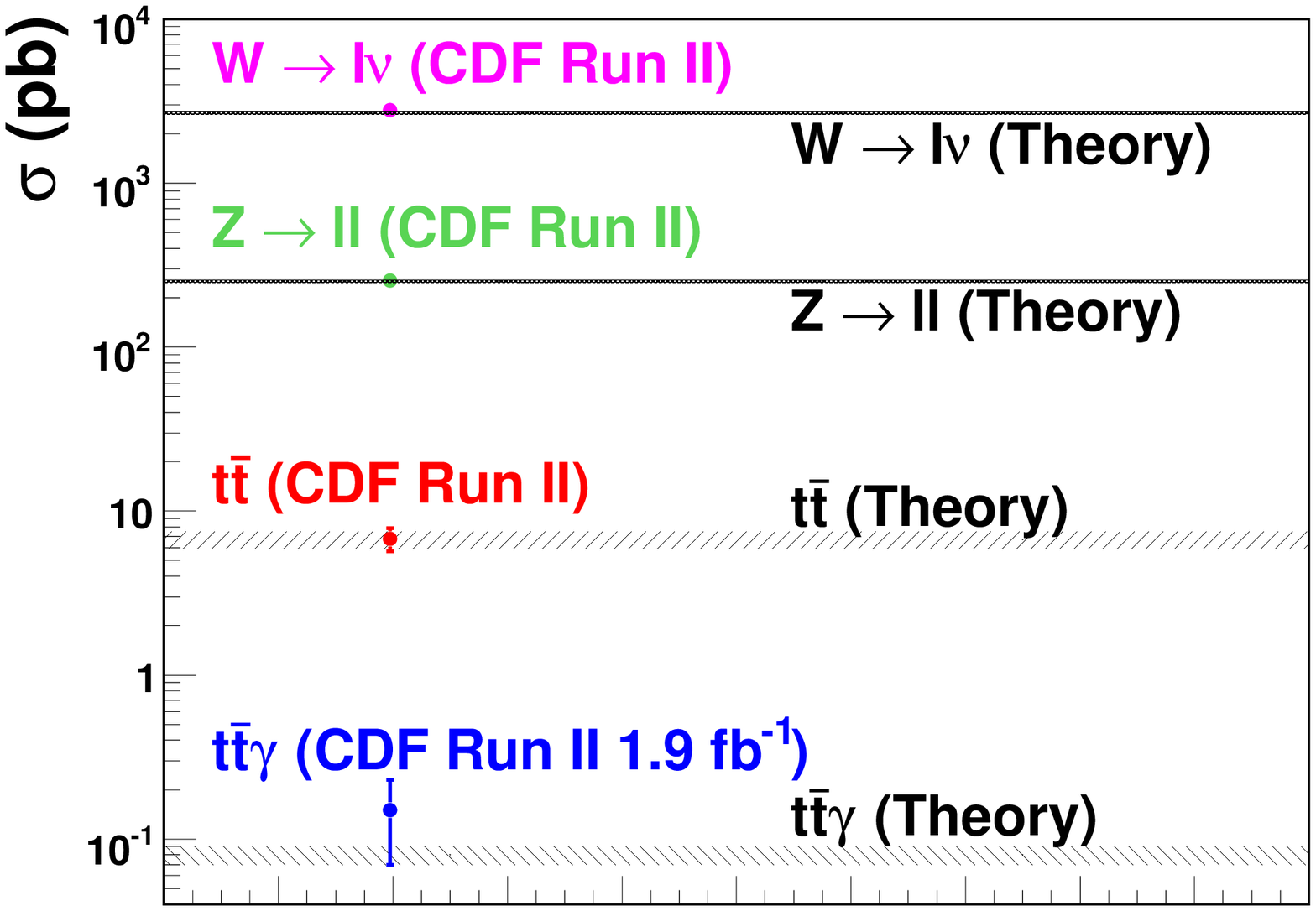}
\end{center}
\vspace*{-0.2in}
\caption
{Estimate of $\sigma_{\ttg}$ compared with SM expectations and other SM
cross sections $\sigma_{W^{\pm}\rrr\ell^{\pm}\nu}$,
$\sigma_{Z\rrr\lplm}$ and $\sigma_{\ttbar}$~\cite{pdg2008}.}
\label{figure5}
\end{figure}

The probability that the backgrounds alone (i.e. assuming that
there is no SM production of the $\ttg$ final state) will produce 16
or more events, is 1\% (2.3 standard deviations). Assuming that the
difference between the non-top background estimate and the number of
observed events is due to $\ttg$ SM production, we estimate the $\ttg$
cross section to be $0.15\pm0.08~\pb$ (see Fig.~\ref{figure5}). An
estimate of the expected semileptonic cross section $\sigma(SM) =
0.080 \pm 0.011~\pb$ is obtained from the LO {\sc madgraph} cross
section of $0.073~\pb$, multiplied by a K-factor 
($\sigma_{NLO}/\sigma_{LO}$) of $1.10 \pm 0.15$~\cite{baur}.
The uncertainty on the cross section is dominated by the statistical
uncertainties associated with the small number of events observed.

In conclusion, we have performed a search for events containing a
lepton, photon, $b$-quark production, and missing $\Et$, a channel
which contains a vector boson and a third-generation quark and is
suppressed in the SM. We find no evidence for non-SM production.  As
an extension of this search we have also performed a search for the SM
process $\ppbar\goes\ttg$, which is predicted to be the dominant
process that produces this signature with at least 3 jets and large
total transverse energy $\Ht$. Here too we find good agreement with
the SM expectations. Although not statistically significant, the
number of observed $\ttg$ events is larger than the SM prediction not
including $\ttg$ production.  Assuming the difference between the
observed number and the predicted non-top-quark SM total is due to top
quark production, we estimate the $\ttg$ cross section to be $0.15 \pm
0.08~\pb$.

\begin{acknowledgments}
We thank the Fermilab staff and the technical staffs of the
participating institutions for their vital contributions. Uli Baur,
Johan Alwall, Fabio Maltoni, Michel Herquet, Zack Sullivan, Stephen
Mrenna, Frank Pietrello, and Tim Stelzer were extraordinarily helpful
with the SM predictions. We thank the Fermilab staff and the technical
staffs of the participating institutions for their vital
contributions.  This work was supported by the U.S. Department of
Energy and National Science Foundation; the Italian Istituto Nazionale
di Fisica Nucleare; the Ministry of Education, Culture, Sports,
Science and Technology of Japan; the Natural Sciences and Engineering
Research Council of Canada; the National Science Council of the
Republic of China; the Swiss National Science Foundation; the
A.P. Sloan Foundation; the Bundesministerium f\"ur Bildung und
Forschung, Germany; the Korean Science and Engineering Foundation and
the Korean Research Foundation; the Science and Technology Facilities
Council and the Royal Society, UK; the Institut National de Physique
Nucleaire et Physique des Particules/CNRS; the Russian Foundation for
Basic Research; the Ministerio de Ciencia e Innovaci\'{o}n, and
Programa Consolider-Ingenio 2010, Spain; the Slovak R\&D Agency; and
the Academy of Finland.
\end{acknowledgments}


\begin{thebibliography}{99}
\bibitem{EtPt} 
Transverse momentum and energy are defined as $\pt = p\sin\theta$ and
$\Et = E\sin\theta$, respectively.
%
Missing $\rm E_T$ ($\metvec$) is defined by $\metvec = -\sum_{i} E_T^i
\hat{n}_i$, where $i$ is the calorimeter tower number for $|\eta| <$ 3.6
(see Ref.~\cite{CDF_coo}), and $\hat{n}_i$ is a unit vector
perpendicular to the beam axis and pointing at the $i^{th}$ tower. We
correct $\metvec$ for jets and muons. We define the magnitude
$\met=|\metvec|$.
%
We use the convention that ``momentum'' refers to $pc$ and ``mass'' to $mc^2$.
\bibitem{CDF_coo} 
The CDF coordinate system of $r$, $\varphi$, and $z$ is cylindrical,
with the $z$-axis along the proton beam. The pseudorapidity is $\eta =
-\ln(\tan(\theta/2))$.
\bibitem{jet_counting}  
Jets that coincide with an identified electron or photon are removed;
each calorimeter cluster is with either a jet, an electron, or a
photon that have mutually exclusive definitions to avoid any
ambiguities.
\bibitem{PhoIso}
The $\Et$ deposited in the calorimeter towers in a cone in
$\eta-\varphi$ space~\cite{CDF_coo} of radius $R=0.4$ around the
photon or lepton position is summed, and the $\Et$ due to the photon
or lepton is subtracted. The remaining $\Et$ is required to be less
than $2.0~\GeV+0.02\times(\Et-20~\GeV)$ for a photon, or less than
10\% of the $\Et$ for electrons or $\pt$ for muons. In addition, for
photons the sum of the $\pt$ of all tracks in the cone must be less
than $2.0~\GeV+0.005\times\Et$.
\bibitem{ElePt}
A high-quality track with $\pt>0.5~\Et$, unless $\Et > 100~\GeV$, in
which case the $\pt$ threshold is set to $20~\GeV$.
\bibitem{hadoem} 
The fraction of electromagnetic energy allowed to leak into the hadron
compartment $\rm E_{had}/E_{em}$ must be less than $\rm
0.055\plus0.00045\times E_{em}~(\GeV)$ for central electrons, less
than 0.05 for electrons in the end-plug calorimeters, less than
max[0.125, $\rm 0.055\plus0.00045\times E_{em}~(\GeV)$] for photons.
\end{thebibliography}
\end{document}